\documentclass[fleqn,usenatbib]{mnras}

\usepackage{newtxtext,newtxmath}
\usepackage[T1]{fontenc}
\usepackage{ae,aecompl}

\usepackage{graphicx}	% Including figure files
\usepackage{amsmath}	% Advanced maths commands
\usepackage{amssymb}	% Extra maths symbols

\title[Asteroseismology of $\mu$ Herculis using SONG data]{Asteroseismic modelling of the subgiant $\mu$ Herculis using SONG\thanks{Based on observations made with the SONG telescopes operated on the Spanish Observatorio del Teide (Tenerife) and at the Chinese Delingha Observatory (Qinghai) by the Aarhus and Copenhagen Universities, by
the Instituto de Astrof\'{i}sica de Canarias and by the National Astronomical Observatories of China.} data: lifting the degeneracy between age and model input parameters}

\author[Tanda Li et al.]{
Tanda Li$^{1,2,3}$\thanks{E-mail: tanda.li@sydney.edu.au},
Timothy R. Bedding$^{1,2}$ ,
Hans Kjeldsen$^{2}$,
Dennis Stello$^{4,2}$,
\newauthor
J\o{}rgen Christensen-Dalsgaard$^{2}$ and 
Licai Deng$^{5}$
\\
$^{1}$Sydney Institute for Astronomy (SIfA), School of Physics, University of Sydney, NSW 2006, Australia\\
$^{2}$Stellar Astrophysics Centre, Department of Physics and Astronomy, Aarhus University, Ny Munkegade 120, DK-8000 Aarhus C, Denmark\\
$^{3}$Key Laboratory of Solar Activity, National Astronomical Observatories, Chinese Academy of Science, Beijing 100012, China\\
$^{4}$School of Physics, University of New South Wales, NSW 2052, Australia\\
$^{5}$Key Laboratory for Optical Astronomy, National Astronomical Observatories, Chinese Academy of Sciences, Beijing 100012, China
}

\date{Accepted XXX. Received YYY; in original form ZZZ}

\pubyear{2015}

\begin{document}
\label{firstpage}
\pagerange{\pageref{firstpage}--\pageref{lastpage}}
\maketitle

% Abstract of the paper
\begin{abstract}
We model the oscillations of the SONG target $\mu$ Herculis to estimate the parameters of the star. The $\ell$ = 1 mixed modes of $\mu$ Her provide strong constraints on stellar properties. The mass and age given by our asteroseismic modelling are 1.10$^{+0.11}_{-0.06}$ M$_{\odot}$ and 7.55$^{+0.96}_{-0.79}$ Gyr. The initial helium abundance is also constrained at around $Y_{\rm{init}}$ = 0.28, suggesting a ratio in the elements enrichment law ($\Delta Y/\Delta Z$) around 1.3, which is closed to the solar value. The mixing-length parameter converges to about 1.7, which is $\sim$ 10\% lower than the solar value and consistent with the results from hydrodynamic simulations.
Our estimates of stellar mass and age agree very well with the previous modelling results with different input physics. Adding asteroseismic information makes these determinations less model-dependent than is typically the case when only surface information is available.
Our studies of the model dependence (mass, initial helium and metallicity fractions, and the mixing length parameter) of the age determination indicate that accurate stellar ages ($\lesssim$ 10\%) can be expected from asteroseismic modelling for stars similar to $\mu$ Her. 
The $\ell$ = 1 bumped modes, which are sensitive to the mean density of the helium core, provide a useful `clock' that provides additional constraints on its age.
\end{abstract}

% Select between one and six entries from the list of approved keywords.
% Don't make up new ones.

%%%%%%%%%%%%%%%%%%%%%%%%%%%%%%%%%%%%%%%%%%%%%%%%%%

%%%%%%%%%%%%%%%%% BODY OF PAPER %%%%%%%%%%%%%%%%%%
\section{Introduction}
Asteroseismic modelling of subgiants can give precise determinations of their global properties (mass, radius, gravity, etc.) \citep{CD95,Carrier05,Brandao2011,Dogan13,Tian15}. 
\citet{Gilliland10}, \citet{Chaplin10} and \citet{Metcalfe10} also raised the exciting possibility that detailed modelling of these stars could provide a very precise determination of their age. Moreover, \citet{Fernandes03} and \citet{Pinheiro10} suggested that the subgiants with stellar oscillations are a good sample for studying the degeneracy of model solutions. They studied $\beta$ Hydri and $\mu$ Herculis and found that their model solutions were insensitive to convection parameters (mixing-length and overshooting parameters). Unlike main-sequence stars, subgiants show mixed modes that provide additional constraints on the structures near the core, which play a key role in stellar evolution. In particular, in the subgiant phase the core sets the temperature of the hydrogen-burning shells above it.The properties of mixed modes (frequency, height, and width) carry information on the deeper layers and place stringent constraints on stellar physics and the global parameters \citep{Benomar2013}.
%however, the helium abundance and stellar mass resulted in a strong degeneracy. 

Bumping of mixed $\ell =1$ modes in subgiants was observed with ground-based telescopes in $\eta$~Boo \citep{Kjeldsen95,Kjeldsen2003,Carrier05} and $\beta$~Hyi \citep{Bedding2007}, and identified as mixed modes by \citet{CD95}. Many more examples have been seen by the space missions CoRoT (Convection, Rotation and planetary Transits space mission) and {\em Kepler}. For example, \citet{DM11} used the $\ell$ = 1 avoided crossing of HD 49385 to constrain the mixing-length and overshooting parameter of its core. The theoretical study of \citet{Benomar2012} found that the coupling strength of the $\ell$ = 1 mixed modes at the subgiant stage was predominantly a function of stellar mass and appears to be independent of the metallicity. 
\citet{Tian15} used the frequency difference between the mixed modes and the nearest p-mode ($d\nu_{m-p}$) as criteria to constrain the stellar models of KIC 6442183 (``Dougal'') and KIC 11137075 (``Zebedee'').
\citet{Ge15} used the mixed modes to constrain the alpha-enhancement of {\em Kepler} subgiant KIC 7976303. 
Moreover, \citet{Bedding2014} suggested that a new asteroseismic p-g diagram, in which the frequencies of the avoided crossings 
%jcd comments: we have not introduced gamma modes. Out of curiosity, can we give references where this diagnostic has actually been used??]
(corresponding to the so-called `$\gamma$ modes', i.e, the pure g modes in the core) are plotted against the large separation of the p modes, could prove to be an instructive way to display results of many stars and to make a first comparison with theoretical models \citep[e.g.,][]{Campante2011}.

\subsection{$\mu$ Herculis}

The star $\mu$ Herculis (HD 161797, HR 6623, HIP 86974) is a nearby G5 IV subgiant star, at a distance of 8.3 pc \citep{Van07}. Its brightness (V = 3.42) makes
the star one of the most studied solar-type stars with ground observations.  
A good summary of the literature values of its effective temperature, gravity, and metallicity can be found in \citet{Soubiran16}. 
Compared with the Sun, it is relatively metal-rich (literature values of [Fe/H] range from +0.04 to +0.3) and slightly cooler ($T_{\rm{eff}}$ = 5397 -- 5650 K), with a $\log g$ from 3.7 to 4.1. 
The parallax (120.33$\pm$0.16 mas from Hipparcos and 119.11$\pm$0.48 mas from Gaia DR2) and the angular diameter given by the Centre for High Angular Resolution Astronomy Array (CHARA) \citep{ten05} also provide measurements of its luminosity and radius. 

$\mu$ Her is a slightly evolved solar-like oscillator with $\nu _{\rm{max}}$ = 1216 $\pm$ 11 $\mu {\rm Hz}$ and $\Delta \nu$ = 64.2 $\pm$ 0.2 $\mu {\rm Hz}$. 
\citep{G17}. An avoided crossing due to coupling between $\ell$ = 1 acoustic and gravity modes appears at $\sim$ 800 $\mu {\rm Hz}$. \citet{Bonanno2008} first detected the solar-like oscillations of the star and extracted 15 individual frequencies based on seven-nights of time-series data observed by the Italian 3.6 m telescope TNG on La Palma. The observed frequencies have been modelled by \citet{YM10} to determine the mass and age. The SONG (Stellar Observations Network Group) telescope at Observatorio del Teide on Tenerife started to observe $\mu$ Her in 2014. With the radial-velocity data collected in the first two years, a total of 49 oscillation modes with $\ell$ value from 0 to 3 were detected \citep{G17}. The asteroseismic modelling for the data implied a stellar mass of 1.11$\pm$0.01 $M_{\odot}$ and an age of 7.8$\pm 0.4$ Gyr. \citet{Kjeldsen18} recently updated the observed frequencies based on the full-set data observed from 2014 to 2017. The data include four-seasons data from Observatorio del Teide on Tenerife, and one-season (2017) data from the second SONG node at Observatory Delingha, China. A total of 50 individual frequencies were extracted with improved signal-to-noise ratios. Comparing with the earlier detections
\citep{G17}, three new high-frequency modes for
$\ell$ = 0, 1, and 2 from 1567 to 1601 $\mu \rm{Hz}$, as well as three new
$\ell$ = 3 modes, were extracted from the new data. However,
three low-frequency modes from 667 to 703 $\mu \rm{Hz}$ and two
high-frequency modes from 1635 to 1670 $\mu \rm{Hz}$ were removed
due to their low S/N.

In this work, we aim to extend the modelling of $\mu$ Her beyond previous efforts with an emphasis on studying the degeneracy of the model solutions (due to uncertainty in the input physics) in order to obtain reliable stellar parameters, particularly mass and age.

\section{Observed Constraints and Theoretical models}%\label{sec:maths}

\subsection{Observed Constraints}

\begin{figure}
	% To include a figure from a file named example.*
	% Allowable file formats are eps or ps if compiling using latex
	% or pdf, png, jpg if compiling using pdflatex
	\includegraphics[width = \columnwidth]{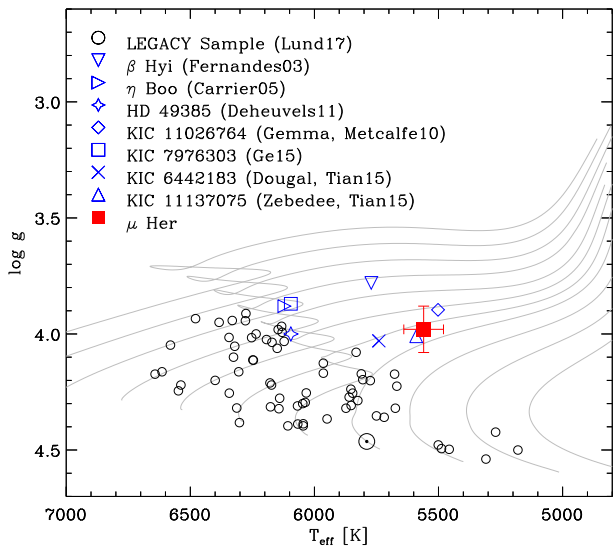}
    \caption{Stars in the $\log g$-$T_{\rm{eff}}$ diagram. Open circles indicate the {\em Kepler} LEGACY sample; open blue symbols are the subgiants which are similar to $\mu$ Herc and have been modelled in the literature; the red filled square shows the location of $\mu$ Her; and $\odot$ represents the Sun. The grey lines are theoretical evolutionary tracks with masses from 0.8 to 1.8 $M_{\odot}$.      
    \label{fig:hr}}
\end{figure}

In this work, we adopted the same non-seismic observational constraints as those used by \citet{G17}, to allow a proper comparison with their modelling results. The
values of $T_{\rm{eff}}$, $\log g$, and [Fe/H] were given by \citet{J15} and \citet{G17} determined their uncertainties. The luminosity and the radius were estimated by \citet{G17} based on the Hipparcos parallax and the angular diameter.
Note that the Gaia DR2 parallax (119$\pm$0.48 mas) has greater uncertainty than the Hipparcos parallax (120.33$\pm$0.16 mas), so we did not use it.
And we adopted the oscillation frequencies recently updated by \citet{Kjeldsen18} for the asteroseismic modelling. 
A summary of the properties can be found in Table 1. Figure \ref{fig:hr} shows the location of $\mu$ Her in the $\log g$-$T_{\rm{eff}}$ diagram compared with the {\em Kepler} LEGACY sample of main-sequence stars,
plus the subgiants similar to $\mu$ Her that have been modelled in the literature. Among these, $\mu$ Her is very similar to the {\em Kepler} star KIC 11137075 (``Zebedee"), both in spectroscopic and asteroseismic features \citep[Fig. 4]{Tian15}. 

% Example table
\begin{table}
	\centering
	\caption{Properties of $\mu$ Her}
    \resizebox{\columnwidth}{!}{
	\label{tab:obs}
	\begin{tabular}{lccr} % four columns, alignment for each
		\hline
		 Parameter & Value & Ref.\\
		\hline
		$T_{\rm{eff}}$ [K]& 5560(80) & \citet{J15,G17} \\
		$\rm{[Fe/H]}$ [dex] & 0.28(7) & \citet{J15,G17} \\
		$\log g$ [dex] & 3.98(10) & \citet{J15,G17} \\
        $R$/R$_{\odot}$ & 1.73(2)& \citet{G17}\\
        $L$/L$_{\odot}$ & 2.54(8) &\citet{G17} \\ 
		\hline
	\end{tabular}
    }
\end{table}

%\begin{table}
%\label{tab:oscillation}
% \centering
%  \caption{Oscillation Frequency Observed by SONG}
%\[%\resizebox{\columnwidth}{!}{\]
%  \begin{tabular}{cccc}
%  \hline
%  \multicolumn{4}{c}{Fequency [$\mu$Hz]}\\
%   $l$ = 0 & $l$ = 1 & $l$ = 2 & $l$ = 3 \\
%  \hline
% 737.34(35) &     *765.66(26)  & 732.19(32) &     -          \\
% 801.28(28)	&     *827.88(29)  &     795.88(33) &     -          \\
% 864.80(21)	&     *904.34(17)  &     858.28(28) &     -          \\
% 928.13(23)	&     *958.76(17)  &     924.14(28) &     -          \\
% 991.39(19)	&    1020.73(14)  &     987.01(22) &    1015.16(34)  \\
%1054.52(16)	&    1083.70(16)  &    1049.50(19) &    1077.21(37)  \\
%1118.99(15)	&    1147.43(11)  &    1113.22(15) &    1140.15(22)  \\
%1183.29(15)	&    1211.14(10)  &    1178.78(13) &    1204.87(27)  \\
%1247.77(13)	&    1275.09(10)  &    1242.85(13) &    1268.86(19)  \\
%1311.76(13)	&    1339.49(14)  &    1307.05(18) &    1333.74(27)  \\
%1376.56(19)	&    1403.82(16)  &    1371.37(20) &    1397.75(33)  \\
%1442.57(18)	&    1469.24(23)  &    1436.50(21) &    1463.84(29)  \\
%1507.26(24)	&    1534.70(22)  &    1501.99(28) &      -          \\
%1573.54(31)	&    1600.02(23)  &    1568.15(29) &      -          \\
%   \hline
%\multicolumn{4}{l}{Notes: p-g mixed modes are marked by `*'.}
%\end{tabular}
%}
%\end{table}

\subsection{Stellar Models and Input Physics}
In this work, we used Modules for Experiments in Stellar Astrophysics
(MESA, version 8118) to compute stellar evolutionary tracks and generate
structural models. MESA is an open-source stellar evolution package
that is undergoing active development. Detailed descriptions
can be found in \citet{Paxton11,Paxton13, Paxton15}.

We adopted the solar chemical mixture [$(Z/X)_{\odot}$ = 0.0181]
provided by \citet{ASP09}.
The initial chemical composition was calculated by: 
\begin{equation}
\log (Z_{\rm{init}}/X_{\rm{init}}) = \log (Z/X)_{\odot} + \rm{[Fe/H]}_{\rm{init}}  \\
\end{equation}
We used the MESA $\rho-T$ tables based on the 2005
update of OPAL EOS tables \citep{EOS} and OPAL opacity for the
solar composition of \citet{ASP09} supplemented by the
low-temperature opacity \citep{op05}. The mixing
length theory of convection was implemented, where 
$\alpha_{\rm MLT} = \ell_{\rm MLT}/H_p$ is the mixing-length parameter for modulating
convection. Convective overshooting of the core was set as described by
\citet[Section 5.2]{Paxton11} and the overshooting mixing diffusion
coefficient was
\begin{equation}
{D_{\rm{OV}}} = {D_{\rm{conv,0}}}\exp \left( - \frac{{2z}}{{f_{\rm{ov}}{H
_{p}}}}\right).
\end{equation}
Here, $D_{\rm{conv,0}}$ is the diffusion coefficient from the mixing-length theory at a
user-defined location near the Schwarzschild boundary, $z$ is the
distance in the radiation layer away from the location, and $f_{\rm{ov}}$ is a
free parameter to change the overshooting scale. We
adopted a fixed $f_{\rm{ov}}$ at 0.018 in the grid computation. 
The other free parameter in the MESA model for the exponential overshooting
is $f_{\rm 0}$, which defines a reference point at which the switch from convection to overshooting occurs, and we set $f_{\rm 0}$ = 0.5$f_{\rm{ov}}$. 
The photosphere tables were used as the set of boundary
conditions for modelling the atmosphere. Atomic diffusion of helium and 
heavy elements was also taken into account. MESA calculates particle diffusion and gravitational settling by solving Burger's equations using the method
and diffusion coefficients of \citet{Thoul94}. 
%We considered 8 classes of species (H1, He3, He4, C12, N14, O16, Ne20, and Mg24) We considered 8 classes of species (H1, He3, He4, C12, N14, O16, Ne20, and Mg24)
We considered eight elements (${}^1{\rm H}, {}^3{\rm He}, {}^4{\rm He}, {}^{12}{\rm C}, {}^{14}{\rm N}, {}^{16}{\rm O}, {}^{20}{\rm Ne}$, and ${}^{24}{\rm Mg}$)
for diffusion calculations, and had the charge calculated by the MESA ionization module, which estimates the typical ionic charge as a function of $T$, $\rho$, and free electrons per nucleon from \citet{Paquette1986}.

The MESA astero extension's 'grid search' 
function was used to generate the grid and find the well-fitting models.
We used the GYRE code in the MESA package to calculate 
the stellar oscillations. The `grid search' function is an efficient
method for searching the best-fitting models in a given parameter space. 
During the model computation, the `grid search' function checks the agreement of non-seismic parameters in real time and calculates the oscillation frequencies when the $\chi ^2_{\rm{non-seismo}}$ value falls below a user-defined threshold. 
We adopted five non-seismic constraints ($T_{\rm{eff}}$, $L$, $R$, $\log g$, and [Fe/H]) in the computation, and the threshold of the average $\chi ^2$ was given as 4.1, for which the cumulative probability of the $\chi ^2$ distribution for five degrees of freedom is 0.999.
When the model passes the non-seismic threshold, MESA computes the radial modes ($\ell$ = 0) and compares the results with the observed frequencies using a least-square test. Here, a user-specified seismic $\chi^2$ is required for determining whether the computations of non-radial modes ($\ell$ > 0) should be carried out. We used a seismic threshold of 30. 
%Due to the two thresholds, MESA does not output the best-fitting models for every evolutionary track. 
An advantage of this method is that the evolutionary time step is adjusted automatically based on the goodness (decreases with $\chi ^2$) of fit to avoid missing the best model. This is very useful for the modelling of subgiant stars, whose internal structure changes rapidly with the age. However, this means that the time steps for different evolutionary tracks vary significantly, depending on their fitting results. Hence, the grid models are far from uniformly spaced in the age. For this reason, only the best-fitting model on each evolutionary track was adopted in the statically studies. For further reference, the MESA inlist used in this work can be found at \url{ http://www.physics.usyd.edu.au/~tali8156
/MESA_inlist/}.

\subsection{Fitting Method}
The fitting method adopted in the MESA `grid search' function is the least-$\chi ^2$ test. The non-seismic $\chi ^2$ (given by $T_{\rm{eff}}$, $\log g$, [Fe/H], $L$, and $R$ in this work) and the 
seismic $\chi ^2$ (given by 50 individual frequencies) are calculated separately with the equations given below,
\begin{equation}
\chi _{\rm{non-seismo}}^2 {\rm{ = }} -\frac{{{1}}}{5}\sum\limits_{i = 1}^5
\left({\frac{{{{(\it{x_i} - \it{u_i})}^2}}}{\it{\sigma _i^2}}}\right),
\end{equation}
and
\begin{equation}
\chi _{\rm{seismo}}^2 {\rm{ = }} -\frac{{{1}}}{50}\sum\limits_{i = 1}^{50}
\left({\frac{{{{(\it{\nu _{mod,i}} - \it{\nu _{obs,i}})}^2}}}{\it{\sigma _{obs,i}^2}}}\right), 
\end{equation}
where $\it{x_i}$ is the i-th theoretical global parameter, $\it{u_i}$ and $\it{\sigma _i}$ is the observed value and its uncertainty, $\it{\nu _{mod,i}}$ is the model frequency, $\it{\nu _{obs,i}}$ and $\sigma _{obs,i}$ is observation and its uncertainty.
The final $\chi ^2$ is the weighted sum of both with user-specified weights. It hence can be described as  
\begin{equation}
\chi ^2 {\rm{ = }} w_{1}\chi _{\rm{non-seismo}}^2 + w_{2}\chi _{\rm{seismo}}^2,
\end{equation}
where $w_{1}$ and $w_{2}$ ($w_{1} + w_{2} = 1.0$) are the weights for the non-seismic and seismic constraints.

\subsection{The Surface Correction} 
The systematic offset between the observed and theoretical oscillation frequencies, known as the surface term, is caused by the improper modelling of the surface of stars.
We used the combined expression described by \citet{ball14}
for correcting the surface term. 
The correction formula is a combination of inverse 
and cubic terms based on the discussion of potential asymptotic forms
for frequency shifts \citep{Gough90}, and it is written as     
\begin{equation}
\delta \nu  = ({a_{-1}}{(\nu /{\nu _{ac}})^{-1}} + {a_3}{(\nu
/{\nu _{ac}})^3})/I,
\end{equation}
%Refer back to them as e.g. equation~(\ref{eq:quadratic}).
where $I$ is the normalised mode inertia and ${a_{-1}}$
and ${a_3}$ are coefficients adjusted to obtain the best frequency 
correction($\delta \nu$). The method to determine these two
coefficients for a given set of observed and model frequencies was described
by \citet{ball14}. 
According to the frequency offsets found for the Sun
and other well-studied solar-like stars \citep{ball14,kje08},
$\delta \nu$ increases with frequency and is also modulated by the mode inertia.
In Eq. 4, ${\nu _{\rm{ac}}}$ is the acoustic cut-off
frequency, which is derived from the scaling relation \citep{Brown91}:
\begin{equation}
\frac{\nu _{\rm{ac}}}{\nu _{\rm{ac, \odot}}} \approx \frac{g}{{{g_
\odot}}}\left(\frac{{{T_{\rm{eff}}}}}{{{T_{\rm{eff, \odot }}}}}\right)^{-1/2}.
\end{equation}
Here we take $\nu _{\rm{ac, \odot}}$ = 5000 $\mu {\rm Hz}$ following \citet{ball14}. Solar values of effective temperature and surface gravity are $T_{\rm{eff},\odot}$ = 5777 K and $\log g$$_{\odot}= 4.44$ \citep{Cox2000}.

\section{The Asteroseismic Modelling}

\subsection{Test Computation}

We first did a test computation within a small grid for two purposes: one was estimating the residual of the surface term after the correction; the other was determining the weights for the non-seismic and seismic $\chi^2$ described in Eq. 6. The test grid was computed with the input mass ranging from 0.9 to 1.3 $\rm M_{\odot}$ (in steps of 0.02 $\rm M_{\odot}$), initial metallicity from 0.25 to 0.45 (increasing by 0.10), but fixed initial helium abundance (0.30) and mixing-length parameter (1.8). Equal weights for non-seismic and seismic $\chi^2$ for the `grid search' function were used.

\begin{figure}
	% To include a figure from a file named example.*
	% Allowable file formats are eps or ps if compiling using latex
	% or pdf, png, jpg if compiling using pdflatex
	\includegraphics[scale = 0.8]{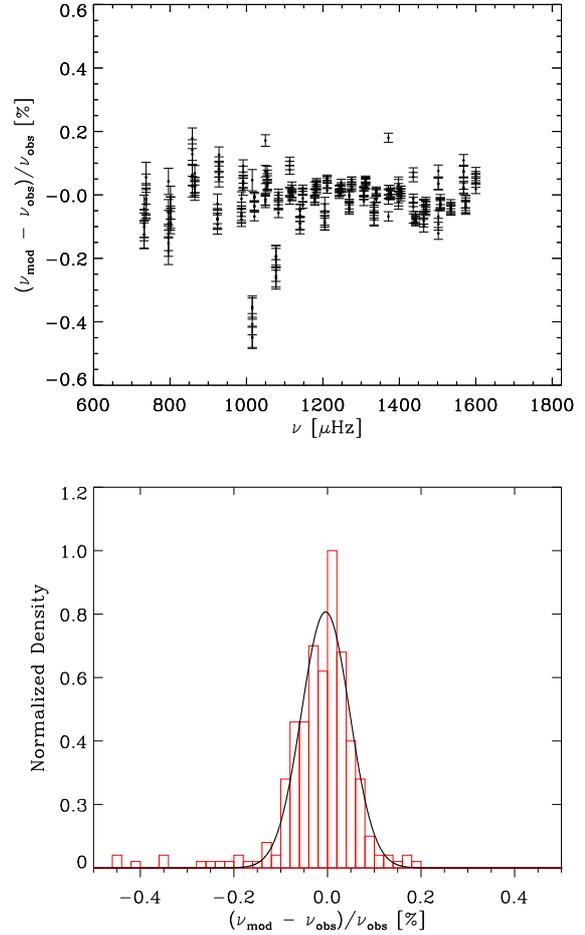}
    \caption{The systematic uncertainty between observed and theoretical
    oscillation frequencies of p/p-dominated modes. Top: frequency differences  
    between the observations and the top-5 models in the test grid as a    
    function of frequency.
    Bottom: the density distribution of the frequency differences. 
    \label{fig:dif}}
\end{figure}

Note that the surface correction methods were developed for removing the offsets from model frequencies. These methods work fairly well on the Sun and most solar-like oscillators; however, the systematic offsets cannot be entirely removed in every single case. That is to say, a certain level of the residual could still exist between model and observed frequencies after the surface correction. Based on the previous studies \citep{ball14,ball17}, the residual of the surface term could be comparable to or even larger than the observed uncertainties. If this is not considered in the fitting procedure, it will lead to modes with different levels of offset having very different weights. 
The surface term varies with mode frequency and inertia. For the case of $\mu$ Her, the $\ell$ = 1 bumped modes at relatively low frequencies should display very small surface effects comparing with the acoustic modes.
This means the systematic uncertainties of p- and p-dominated modes are larger than for the bumped modes. 
The fitting procedure may therefore give higher weight to p- and p-dominated modes than to 
the bumped modes.
Hence, estimating the residual (the systematic uncertainties from models) after the surface correction is necessary. A proper way to do this estimation could be comparing
the frequency differences between the observations and the best-fitting models.
After visual inspection of the \'echelle diagram, we took the top 5 models in the test grid and calculated the frequency differences between the observed frequencies and corrected theoretical frequencies (only for p/p-dominated modes). The results are shown in the top in Figure \ref{fig:dif} as a function of observed frequencies. Obvious scatter appears from -0.1$\%$ to +0.1$\%$. The histogram of the frequency differences given at the bottom suggests a systematic uncertainty of $\pm$0.05$\%$ of the frequencies, 
corresponding to $\pm$0.56 $\mu {\rm Hz}$ at $\nu _{\rm{max}}$. Compared with the observational uncertainties ($\sim$ 0.2 $\mu {\rm Hz}$), this residual is certainly too large to be neglected. Hence, we considered this systematic uncertainty in addition to the observed uncertainty in the following fitting process.

We next determined relative weights ($w_{1}$ and $w_{2}$ in Eq. 5) of the non-seismic and seismic constraints. Based on the previous results, we considered extra 0.05$\%$ in the uncertainties of all p/p-dominated modes and recalculated $\chi _{\rm{seismo}}^2$. We then visually inspected theoretical models in the \'{e}chelle diagram and found that the models having reasonable fitting to the data were below $\chi _{\rm{seismo}}^2$ $\sim$ 40.  
Because the $\chi _{\rm{non-seismo}}^2$ of all computed models was below 4.1 (the threshold for seismic computations), $\chi _{\rm{seismo}}^2$ would dominate the final $\chi ^2$ if equal weights were used. Considering the accuracy of the observations, seismic data are certainly the primary constraints; however, we still need the non-seismic data to provide secondary constraints. With the specified weights, we wanted $\chi _{\rm{seismo}}^2$ to dominate the final $\chi ^2$ if the models were not asteroseismically good and $\chi _{\rm{non-seismo}}^2$ to kick in when the models showed good fitting with seismic frequencies. For this purpose, we determined the weights as 0.9 and 0.1 for non-seismic and seismic $\chi^2$.

\subsection{Grid Computation}
\label{subsec:grid}
\begin{table}
 \centering
  \caption{Input Parameters of the MESA `Grid Search'}
  \label{tab:grid}
%\resizebox{\columnwidth}{!}{
  \begin{tabular}{ccc}
  \hline
    Input parameters& range & increment \\
  \hline
   M/M$_{\odot}$ &0.90 -- 1.30 & 0.02 \\
   $\rm{[Fe/H]_{init}}$ & 0.25 -- 0.45 & 0.10 \\
   $Y_{\rm{init}}$ & 0.26 -- 0.36&0.02\\
   $\alpha$ & 1.6 -- 2.1 & 0.1\\
   $f_{\rm{ov}}$/$f_{\rm{0}}$&0.018/0.009&-\\
   \hline
   Non-seismic constraints &  \multicolumn{2}{c}{$T_{\rm{eff}}$, $L$, $R$, $\log g$, [Fe/H]} \\ 
   Seismic constraints & \multicolumn{2}{c}{Individual modes for $\ell$ = 0-3} \\
   Weights in Eq. 5 & \multicolumn{2}{c}{$w_1$ = 0.9, $w_2$ = 0.1} \\
   \hline
\end{tabular}
%}
\end{table}

Based on the results obtained in the test computation, we 
considered additional uncertainties (0.05\%) for the p/p-dominated modes and
used $w_1 = 0.9$ and $w_2 = 0.1$ as the weights of $\chi^2$ in Eq. 5.
The grid computation had four independent input parameters:
mass ($M$), initial metallicity ([Fe/H]$_{\rm{init}}$), initial
abundance of helium ($Y_{\rm{init}}$), and the mixing-length parameter ($\alpha$). The ranges and the grid steps of these input parameters are listed in Table \ref{tab:grid}. The `grid search' function finally outputted 263 models after the computations. We then compared the models with the observations in the \'echelle diagram. As we found from test computations, the $\chi^2_{\rm{seismo}}$ values for reasonable-fitting models were below $\sim$ 40 and those of well-fitting models were below $\sim$ 25. 

In Table \ref{tab:topmodels}, the input parameters of the models with $\chi^2_{\rm{seismo}}$ $\lesssim$ 40 and $\chi^2$ $\lesssim$ 5.0 are summarised.
It can be seen that our best-fitting models spread in the mass range from 1.08 to 1.26 M$_{\odot}$ and the age range from 6.8 to 8.6 Gyr.
Models with larger mass tend to have lower initial helium but higher initial metallicity and mixing-length parameter. 
The spread across our best-fitting models caused by three free model parameters ($Y_{\rm{init}}$, [Fe/H]$_{\rm{init}}$, and $\alpha_{\rm{MLT}}$) are $\sim$15\% in stellar mass and $\sim$10\% in age. Compared with the strong mass dependence of the age in main-sequence stars, the mass-age relation in $\mu$ Her looks to be rather weak. We discuss this in Section ~\ref{sec:degeneracy}. 

\begin{figure*}
	% To include a figure from a file named example.*
	% Allowable file formats are eps or ps if compiling using latex
	% or pdf, png, jpg if compiling using pdflatex
	\includegraphics[width = \textwidth]{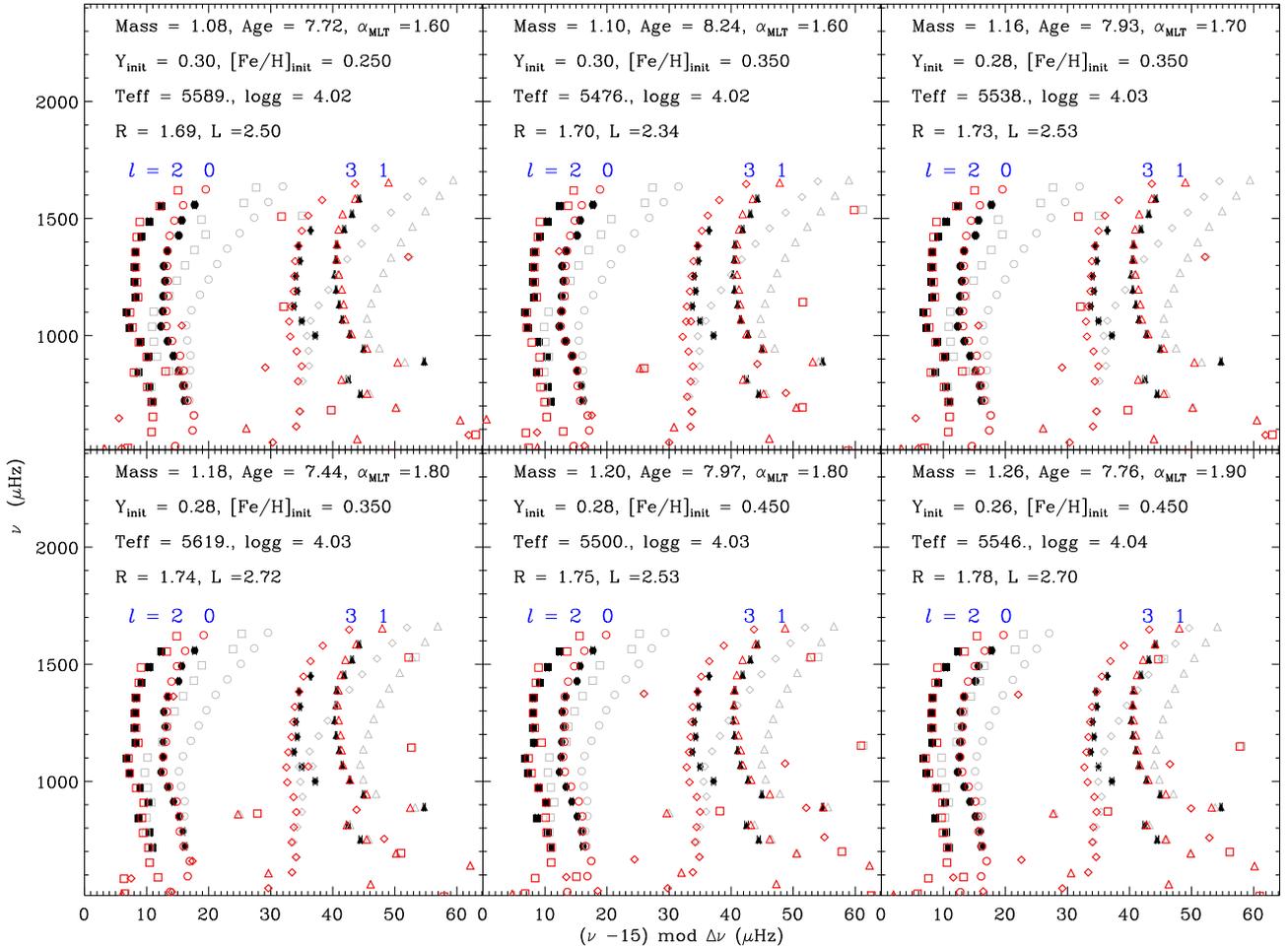}
    \caption{The \'echelle diagram of six of the best-fitting models for $\mu$ Her. Filled black symbols indicate observed frequencies; open grey and red circles ($\ell$ = 0), triangles ($\ell$ = 1), squares ($\ell$ = 2), and domains (($\ell$ = 3)) are model frequencies before and after the surface correction. The input parameters and stellar parameters of the model are given on the top.}
    \label{fig:best_model}
\end{figure*}

From the models in Table \ref{tab:grid}, we selected six with different stellar masses and show their \'echelle diagrams in Figure \ref{fig:best_model}. A few points can be seen by comparing model and observed frequencies. First, good agreement is obtained for most of the radial modes and the p-like modes for $\ell$ = 1 except those with high frequencies. However, these modes have lower S/N (hence larger uncertainty) and are more strongly affected by the surface term than other modes. Second, small bumping can be seen on some of the observed $\ell$ = 2 modes (below 1100 $\mu$Hz) which the models fit fairly well. Third, the two observed $\ell$ = 3 modes from 1000 to 1100 $\mu$Hz could be wrong, because the weak coupling leads to large mode inertia for the $\ell$ = 3 bumped modes which makes them hard to detect. 

Finally, the models show reasonable fits for the $\ell$ = 1 bumped modes (from 700 to 1000 $\mu$Hz); however, none of the models fit all bumped modes. The time step adopted in the computation is probably not the reason for this mismatch, because the bumped modes for $\ell$ = 2, which are more sensitive to the time steps than those for $\ell$ = 1, are well fitted by the model. 
The p-g coupling is strong for $\ell$ = 1 modes but weak for $\ell$ = 2 modes, and it means that the $\ell$ = 1 bumped modes are more sensitive to the core than the $\ell$ = 2 modes. The reason for this imperfect fit is probably that our modelling of the core region is not perfect.

Overshooting, which is not varied in this work, is one of the physical mechanisms that could change the size of the core. Some details about the change in core size and shape with different types of the overshooting were discussed by \citet{Pedersen18}. For the case of $\mu$ Her, which may have had a convective core during the main-sequence stage, the properties of the core that we model would depend on the way overshooting is implemented. This in turn would lead to different core structures during the subgiant phase.
Extra-mixing (caused by rotation, magnetic fields, and internal gravity wave, etc.), which we did not include, could also affect the core and hence the evolution. Extra-mixing could reduce the chemical gradient at the edge of the core and also transport more fuel into the core. Hence models with extra-mixing tend to have a smoother chemical profile at the boundary of the core as well as a longer main-sequence lifetime. Discussions about the effects of the extra-mixing on core structures for the stars within a similar mass range to $\mu$ Her can be found in \citet{ltd2014}. However, one point should be noted: overshooting and extra-mixing can also change the global stellar properties and hence affect the p modes. It is therefore not straightforward to estimate how the modelling results could be affected by a change in the mixing in the framework presented here. Further study would be required to analyse these additional effects.     

\begin{table*}
 \centering
  \caption{Input Parameters of the Models with $\chi ^2$ $\lesssim$ 5.0}
  \label{tab:topmodels}
%\resizebox{\columnwidth}{!}{
  \begin{tabular}{ccccccccc}
  \hline
   No.& Mass & Age& $\alpha$ & $Y_{\rm{init}}$ &[Fe/H]$_{\rm{init}}$ & $\chi ^2 _{\rm{non-seismo}}$ & $\chi ^2 _{\rm{Seismo}}$ &$\chi ^2$\\
  \hline
   1&1.20&       7.97&      1.8&      0.28&      0.45&    0.9&      15.8   &2.44 \\    
   2&1.18&       7.44&      1.8&      0.28&      0.35&    1.2&      15.1   &2.63 \\      
   3&1.16&       7.93&      1.7&      0.28&      0.35&    0.1&      28.3   & 2.89\\     
   4&1.10&       8.24&      1.6&      0.30&      0.35&    1.9&      14.5   &3.19 \\     
   5&1.10&       6.78&      1.8&      0.34&      0.45&    1.5&      22.5   &3.62\\     
   6&1.26&       7.76&      1.9&     0.26&       0.45&    2.9&     12.1   &3.83\\
   7&1.08&       7.26&      1.7&      0.34&      0.45&    1.5&      26.2   &4.00\\      
   8&1.10&       8.43&      1.7&      0.28&      0.25&    1.1&      31.3   &4.15\\  
   9&1.18&       8.49&      1.7&      0.28&      0.45&    2.3&      20.6   &4.15\\      
  10&1.24&       8.24&      1.8&      0.26&      0.45&    1.8&      25.3   &4.17\\ 
  11&1.08&       7.72&      1.6&      0.30&      0.25&    1.4&      29.7  &4.18      \\
  12&1.12&       7.70&      1.7&      0.30&      0.35&	  0.2&      43.2  &4.51\\       
  13&1.10&       7.21&      1.7&      0.30&      0.25&    2.0&      27.7   &4.6\\      
  14&1.14       &8.63&      1.6&      0.26&      0.25    &1.1      &39.9   &4.9\\     
\hline
\end{tabular}
%}
\end{table*}

\subsection{Mass, Age, Helium, and the Mixing-Length Parameter}
\label{subsec:bayesian}
We adopted a Bayesian method to determine the stellar properties.
The theoretical modelling have five independent variables, namely, mass ($M$), initial abundances of helium ($Y$), the ratio between the contents of heavy elements and hydrogen ($Z/X$), the mixing-length parameter ($\alpha_{\rm MLT}$), and the age ($\tau$), to specify a particular model. 
As mentioned above, our method to calculate the $\chi^2$ introduce two adjusted weights to balance
the non-seismic and seismic constraints. However, the downside is introducing some arbitrariness in the analysis. It is hence difficult to analyse in a proper statistical way. We tested a few different ways to calculate the probability distribution of stellar mass and age and found that the inverse $\chi^2$ provides reasonable distributions for stellar parameters.
Thus, we wrote the probability of a model as
\begin{equation}
\it{p}(\rm{Model|Data}){\rm{ = }} \frac{1}{\chi ^2},
\end{equation}
where Model = \{$M$, $Y$, $Z/X$, $\tau$, $\alpha_{\rm MLT}$\}
and Data = \{$T_{\rm{eff}}$, $\log g$ ,[Fe/H], $R$, $L$, $\nu_{n,l}$\}. 
We assumed a flat prior for all model quantities.  
%It should be noted that because the time steps of different evolutionary tracks are not same, we only used the best-fitting model on each evolutionary tracks in the Bayesian analysis to keep the uniform spacing in the parameter space. 
The probability distribution of a given variable can be calculated by marginalising over all the other quantities. For instance, the probability distribution of stellar mass can be calculated by 
\begin{equation}
p(M){\rm{ = }}\int{p(M, Y, Z/X, \tau, \alpha_{\rm MLT})}dY d(Z/X) d\tau d\alpha_{\rm MLT}. 
\end{equation}

\begin{figure*}
	% To include a figure from a file named example.*
	% Allowable file formats are eps or ps if compiling using latex
	% or pdf, png, jpg if compiling using pdflatex
	\includegraphics[width = 0.8\textwidth]{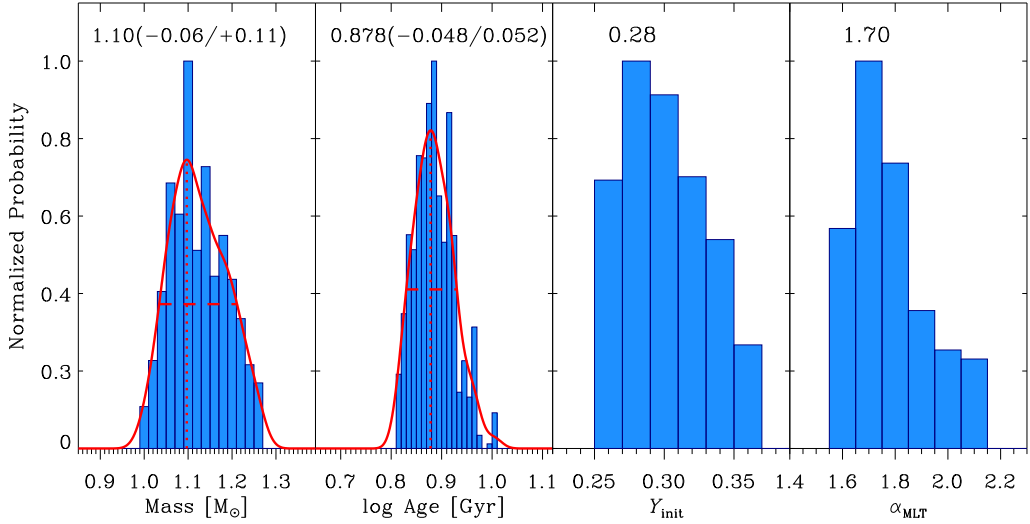}
    \caption{Probability distributions of the mass, the age, initial helium abundances, and the mixing-length parameter for $\mu$ Her. In the first two graphs,  the red solid lines indicate the probability distributions after a Gaussian smoothing, the dotted lines are the highest peaks of the mass and the age, and the dashed lines represent the width at half maximum.
    \label{fig:mass_age_Yinit_alpha}}
\end{figure*}

We first estimated the values of the mass, initial helium abundance, the age, and the mixing-length parameter by using the method mentioned above. 
The probability distributions of the four model parameters are given 
in Figure \ref{fig:mass_age_Yinit_alpha}. 
Even though the mass, the helium, and the mixing-length parameter are all free parameters, we notice that their probability distributions are narrow. This indicates that stellar oscillations seem to be able to lift the degeneracies of these input parameters, particularly between mass and helium/the mixing-length. We will discuss this in Section ~\ref{sec:degeneracy}. 
To estimate the mass and the age of the star, we smoothed the probability distributions (red solid line) and found the highest peak (shown by red dots). The uncertainties were determined by the width at half maximum (red dash). We then estimated the mass and the age of $\mu$ Her as $1.10^{+0.11}_{-0.06}$M$_{\odot}$ and $7.55^{+0.96}_{-0.79}$ Gyr. Our inferred uncertainties of mass and age are larger than those given by \citet{G17}; this is mainly because we considered a relatively large range of the initial helium fraction whereas \citet{G17} anchors it to the heavy element fraction through an enrichment law. 
%\textbf{If we assumed that helium is a function of metals in stars, the initial helium fractions of our model would be in a narrow range and hence our model would give better precisions of the mass and the age.}
%We fitted each probability distribution with Gaussian functions, adopted the centre and 1-$\sigma$ deviation as our estimates. The mass, the age, the initial helium abundance, and the mixing-length parameter of $\mu$ Her were determined as 1.122$\pm$0.073M$_{\odot}$, $7.61^{+0.72}_{-0.66}$ Gyr ($\log \tau$ = 0.882$\pm$0.040), 0.0287$\pm$0.043 and 1.73$\pm$0.14.
%It is interesting that the probability distributions of the mass and initial helium abundance both converge at certain points, although initial helium have strong degeneracy with the mass. The results indicate that the asteroseismic modelling could reduce the degeneracy between the two model parameters for $\mu$ Her. 

The initial helium fraction shows its highest peak in the probability distribution around 0.28 in Fig. \ref{fig:mass_age_Yinit_alpha}, and most of the best-fitting models given in Table \ref{tab:topmodels} have a helium fraction from 0.26 to 0.30, which are slightly higher than the solar value ($\sim$0.27) as given by \citet{ASP09}. 
This result agrees with the typically assumed enrichment law between helium and heavy elements often adopted in Galactic chemical evolution models. 
We then took the helium fraction from our stellar models and the metallicity from observations to determine the ratio between helium and heavy elements in the enrichment law,
\begin{equation}
 {Y_{\rm{init}}} = {Y_0} + \frac{{\Delta Y}}{{\Delta Z}}{Z_{\rm{init}}}. \\
\end{equation}
The ratio (${\Delta Y}/{\Delta Z}$) is an adjusted parameter typically between 1.0 to 3.0. If taking the primordial helium abundance ($Y_0$ = 0.249) determined by \citet{PC16}, then the ratio for the Sun will be 1.50 based on the measurements given by \citet{ASP09}. For the case of $\mu$ Her, we took the observed metallicity ([Fe/H] = 0.35) and helium abundance constrained by our modelling ($Y_{\rm{init}}$ $\sim$ 0.28). Hence the ratio ${\Delta Y}/{\Delta Z}$ of $\mu$ Her is $\sim$1.3, which is similar to the solar value.

Even though our model grid did not cover mixing-length parameter values below 1.6, it clearly converged around $\alpha _{\rm{MLT}}$ = 1.7. And $\alpha _{\rm{MLT}}$ values of almost all the best-fitting models listed in Table \ref{tab:topmodels} fall within 1.6 to 1.9. This is $\sim 10\%$ smaller than the solar value ($\alpha _{\rm{MLT,\odot}}$ =1.9) for the same input physics. This result is consistent with the hydro-simulation based on 3D stellar atmosphere models \citep{Magic15}, which suggested an $\alpha _{\rm{MLT}}$ slightly smaller than solar for stars with similar $T_{\rm{eff}}$ and $\log g$ to $\mu$ Her. 

%however, it is close to the resultsof the LEGACY Sample \citep{legacy17}. 

We summarize our modelling results and compare them with
the estimates given by \citet{G17} in Table \ref{tab:results}. 
The first six lines include the major input physics adopted by \citet{G17} and 
those in this work. The following five lines list the theoretical estimates of the stellar parameters. Although there were many differences in the input physics, 
our estimates of the mass and the age agree with the determinations from \citet{G17} very well.
As the adopted observed constraints were almost the same (only slightly different in observed frequencies), this agreement seems to indicate that the mass and age of $\mu$ Her are not very sensitive to the change of input physics. However, because the sensitivity to the input physics is strongly degenerate and there are also possible differences caused by different numerics between the two codes, this result may need further examination. A similar comparison between the inferred masses and ages with different input physics but using the same code has been done by \citet{DM11}. They studied the CoRoT subgiant HD 49385, which was modelled with two different solar mixtures. Here, the inferred masses and ages resulting from these two grids only differed by 1 $\sigma$ ($\sim$0.02M$_{\odot}$) and 0.5 $\sigma$ ($\sim$0.05Gyr), respectively.

%\subsection{Seismic Stellar Parameters}

%Asteroseismology can also provide independent constraints for stellar parameters. Here we only used oscillation frequencies to calculate the Bayesian probability. The distributions of effective temperature, the luminosity, the radius, and the surface gravity of the star were shown in Figure \ref{fig:seismic_p}. We used Gaussian function to fit each parameter and summarized the results in Table \ref{tab:results}. The seismic parameters have very good agreement with the observational values. Moreover, the precision of $\log g$ has been significantly improved. 

%\begin{figure*}
	% To include a figure from a file named example.*
	% Allowable file formats are eps or ps if compiling using latex
	% or pdf, png, jpg if compiling using pdflatex
	%\includegraphics[scale = 0.8]{seismic_parameters}
    %\caption{Probability distributions of seismic parameters. 
    %The solid lines indicate the Gaussian functions for fitting
    %the distributions.}
    %\label{fig:seismic_p}
%\end{figure*}

\begin{table}
	\centering
	\caption{The Comparison of the Stellar Parameters}
    \resizebox{\columnwidth}{!}{
	\label{tab:results}
	\begin{tabular}{lcc} % four columns, alignment for each
        \hline        
        Input Physics & \citet{G17} & This work \\
        \hline  
        Stellar model & ASTEC + ADIPLS & MESA + GYRE\\
        Solar abundance$^{\rm{a}}$&GS98&AGSS09\\
        Atmosphere&Eddington&MESA Photosphere Table\\
        $Y_{\rm{init}}$ & depending on Z & 0.26 - 0.36 (0.02)\\
        $\alpha_{\odot}$& 1.8& 1.9\\
        $\alpha$&1.5,1.8,2.1&1.6 - 2.1 (0.1)\\        
        Diffusion&No&Yes\\
		\hline
		Model Parameter & \citet{G17} & This work \\
		\hline
        Mass [M$_{\odot}$]&1.11$\pm$0.01& $1.10^{+0.11}_{-0.06}$ \\
        Age [Gyr] &7.3$^{+0.3}_{-0.4}$&$7.55^{+0.96}_{-0.79}$\\
        $Y_{\rm{init}}$&-& $\sim$0.28 \\
        $\Delta Y$/$\Delta Z$ $^{\rm{b}}$&-& $\sim$ 1.3\\
        $\alpha _{\rm{MLT}}$ & - & $\sim$1.7 (0.9 $\alpha_{\odot}$)\\
        \hline
  %      Stellar Parameter & Obs.&Seismic\\
  %      \hline
	%	$T_{\rm{eff}}$ [K]& 5560$\pm$80 & 5580$\pm$95 \\
    % 	$\log g$ [dex] & 3.98$\pm$0.10 & 4.02$\pm$0.01\\
       % $R$/R$_{\odot}$ & 1.73$\pm$0.02& 1.72\pm$0.04\
     %   $L$/L$_{\odot}$ & 2.54$\pm$0.08 &2.56\pm$0.19 \\ 
	%	\hline
	\end{tabular}
    }
    a: GS98: \citet{GS98}, AGSS09: \citet{ASP09}; 
    b: The ratio $\Delta Y$/$\Delta Z$ in Eq. 14. 
    \end{table}

\section{The Degeneracy of Model Solutions in Subgiants}
\label{sec:degeneracy}
%As we mentioned above, the success in constraining both mass and initial helium abundance of the star indicates that the mixed modes help in reducing the degeneracy between the two parameters. Moreover, the agreement between the modelling masses and ages from \citet{G17} and this work suggests less dependence of the modelling results on the input physics.In the grid computations, we varied mass, initial helium,  initial metallicity, and the mixing-length parameters.
The accuracy of the mass and age given by the stellar modelling are usually significantly affected by the changes in input physics and free input parameters.
A comparison between the modelling results of the {\em Kepler} LEGACY sample 
(main-sequence dwarfs) from seven independent pipelines can be found in \citet{legacy17}. 
The typical systematic differences in the estimates of the mass and the age were up to $\sim$0.1M$_{\odot}$ and $\sim$ 1.5 Gyr for these stars, which were much larger than the uncertainties given by each pipeline. The appearance of p-g coupled modes at the subgiant stage provides additional constraints for the properties of the core. 
The avoided crossing in subgiants and its relations to the stellar properties 
have been discussed by \citet{DM11} in detail. Briefly, the p-g coupled modes are sensitive to the mass, the convection parameters, and chemical composition, and hence they can narrow down the dimensions of the model space. In other words, the mixed modes reduce the degeneracy in modelling solutions of subgiants. 
%In this section, we study the degeneracy between the independent input parameters and their effects on the determinations of mass and age.

\subsection{Mass v.s. Age}
\begin{figure}
	% To include a figure from a file named example.*
	% Allowable file formats are eps or ps if compiling using latex
	% or pdf, png, jpg if compiling using pdflatex
	\includegraphics[scale = 0.9]{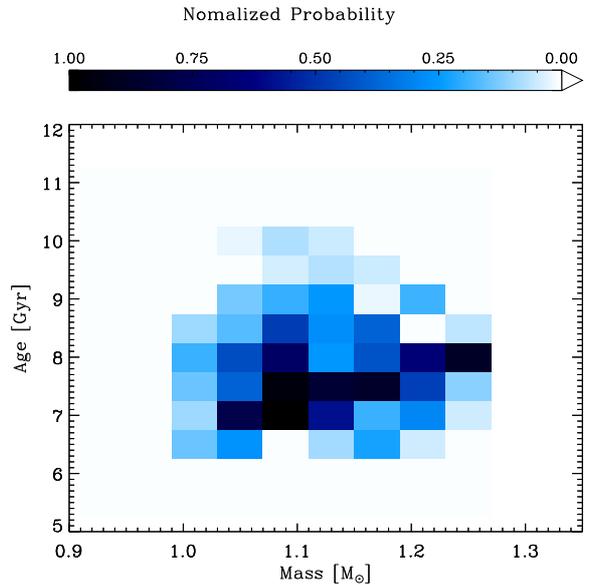}
    \caption{The two-dimension probability distributions of mass and age for $\mu$ Her.}
    \label{fig:mass_age_2d}
\end{figure}

We first present the correlation between mass and age from the grid. Figure \ref{fig:mass_age_2d} is the 2-D probability distribution calculated by the Bayesian method mentioned in Section ~\ref{subsec:bayesian}.   
Due to the dimensions and the large range of the model space, the mass of the models spreads from 1.08 to 1.26 M$_{\odot}$.   
However, the ages corresponding to different masses are of similar values from 6.8 to 8.6 Gyr.
This mass-age relation indicates that the asteroseismic modelling tends to give an age which is only weakly dependent on the stellar mass for $\mu$ Her.    

\subsection{Dependence on the Mixing-length Parameter}

\begin{figure}
    \includegraphics[width=\columnwidth]{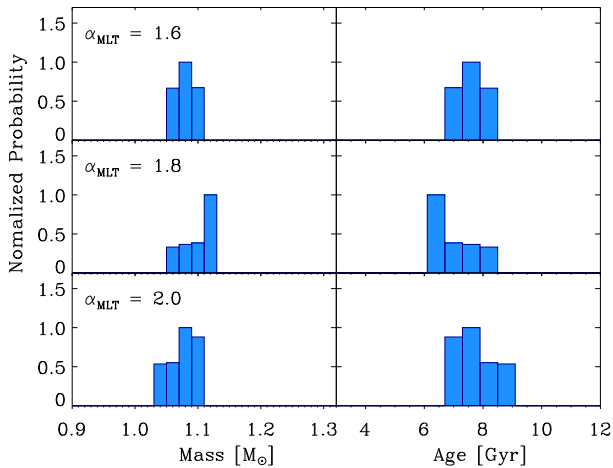}
    \caption{Normalised probability distributions of the mass and age for varying the mixing-length parameter but fixed initial chemical compositions. The initial helium and metallicity are $Y_{\rm{init}}$ = 0.32 and [Fe/H]$_{\rm{init}}$ = 0.35. Model solutions with $\alpha _{\rm{MLT}}$ = 1.6, 1.8, and 2.0 are shown from the top to the bottom.  
    \label{fig:given_alpha}}
\end{figure}

The adjusted mixing-length parameter ($\alpha _{\rm{MLT}}$) is the key factor to determine
the efficiency of energy transport in convective regions in stars. However, previous studies \citep{Fernandes03, Pinheiro10} on $\beta$ Hyi and $\mu$ Her
found no obvious effects from $\alpha _{\rm{MLT}}$ on the determinations of the mass and the age.
Similar results were also obtained by \citet{DM11} when modelling the CoRoT subgiant HD 49385. 
In Figure \ref{fig:given_alpha}, we compare the probability distributions of the mass 
and the age given by the models of $\mu$ Her with different mixing-length parameters but the same initial chemical compositions. Consistent with the previous studies, our modelling mass and age do not show obvious dependence on the mixing-length parameter.

\subsection{Dependence on Initial Helium Abundance}

\begin{figure}
    \includegraphics[width=\columnwidth]{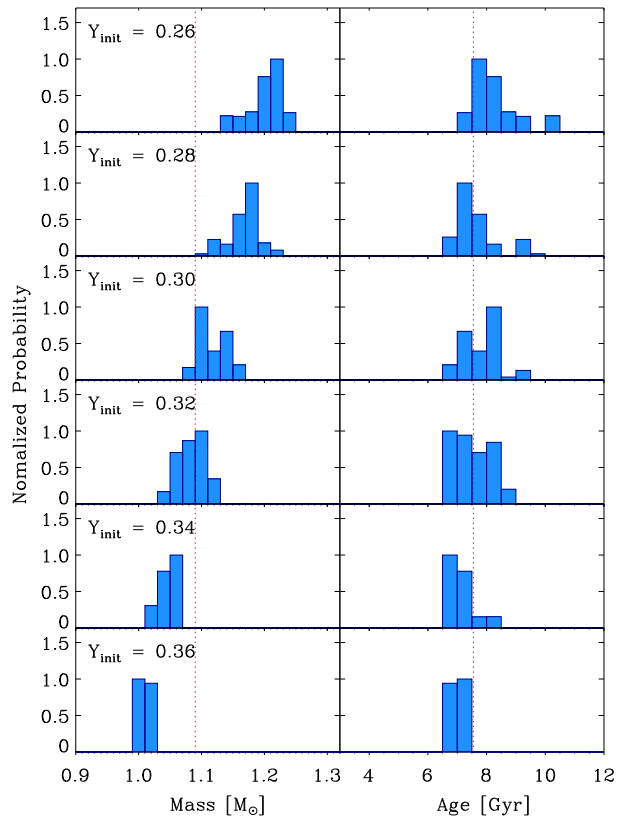}
    \caption{Normalised probability distributions of the mass and age for varying helium abundance but fixed initial metallicity. The initial metallicity is [Fe/H]$_{\rm{init}}$ = 0.35 for all the plots. The helium abundance increases from the top to the bottom.
    \label{fig:given_y}}
\end{figure}

The helium abundance cannot be measured from spectroscopic observations. The lack of knowledge of this parameter significantly affects our understanding of stars. 
The inferred initial helium abundance from stellar modelling has a strong degeneracy with the mass for main-sequence and subgiant stars \citep{Fernandes03, Pinheiro10}. 
%On the contract, it shows small effects on the age \citep[Fig. 8]{Valcarce12}.
To check the effect of helium on asteroseismic modelling solutions, 
we calculated the probability distributions of
mass and age with different initial helium abundances but fixed the mixing-length parameter and initial metallicity. The results are shown in Figure \ref{fig:given_y}. 
%In the left row, a strong correlation between the mass and the helium can be obviously seen. 
The mass changes by $\sim$0.1 $M_{\odot}$ (10\%) and the age changes by $\sim$0.5 Gyr (7\%) when the initial helium decreases from 0.36 to 0.26. 
One interesting point here is that the mass and age change in the same direction, unlike the mass-age relation obtained with varying metallicity or the mixing-length parameter. 
It should be noted that the hydrogen as well as the heavy elements increase when the helium decreases, because we adopted a fixed $Z/X$ but not a fixed $Z$.
%Figures are referred to as e.g. Fig.~\ref{fig:example_figure}, and tables as
%e.g. Table~\ref{tab:example_table}.

\subsection{Dependence on Initial Metallicity}

\begin{figure}
	% To include a figure from a file named example.*
	% Allowable file formats are eps or ps if compiling using latex
	% or pdf, png, jpg if compiling using pdflatex
	\includegraphics[width=\columnwidth]{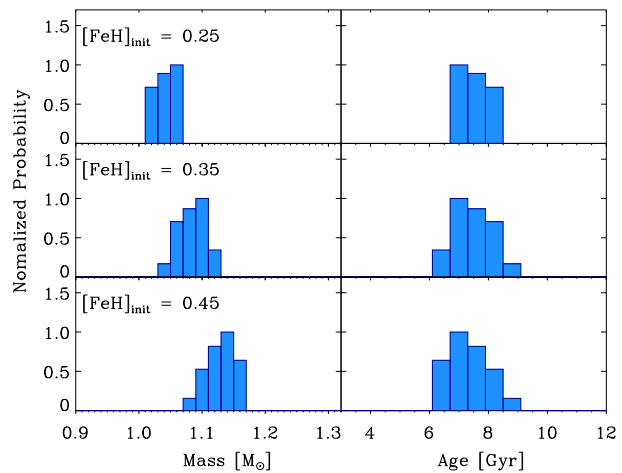}
    \caption{Normalised probability distributions of the mass and age for varying initial metallicity but fixed helium abundance. The initial helium abundance is 0.32 for all the plots. The initial metallicity increases from the top to the bottom. 
    \label{fig:given_z}}
\end{figure}

Although metals only constitute a small fraction of the total mass
of the star, they affect the opacity and hence significantly impact the stellar structure 
and evolution. Hence high-quality measurement of metallicity is usually required for determining stellar parameters.
We calculated the probability distributions of masses and ages for different initial metallicity but the same initial helium abundance as shown in Figure \ref{fig:given_z}. It can be seen that the mass increases by $\sim$5$\%$ while the age hardly changes when the initial metallicity is increased by 0.1 dex. Similar to what we found in Figure \ref{fig:mass_age_2d}, the correlation between stellar mass and age is not very strong here.

\subsection{Accuracy of the age}
Here we discuss the effect of the mixing-length parameter, metallicity, and helium on the accuracy of the age. The mixing-length parameter, as mentioned, does not change the age systematically.
%An offset would be expected due to the lack of observed helium abundances. 
Now, if we assume an enrichment law like Eq. 10 with a $\Delta Y/\Delta Z$ ratio presumably between 1.0 and 2.0 (representative for the solar neighbourhood), we would expect $\mu$ Her to have $Y_{\rm{init}}$ of about 0.27 to 0.30. The change in its age corresponding to this range in helium will be $\sim$2\%. Moreover, the typical observed uncertainty of metallicity ($\sim$0.1 dex) leads an uncertainty in age below 5\%. Hence, good ages ($\lesssim$ 10\%) can be expected from asteroseismology for stars which are similar to $\mu$ Her. 

%\subsubsection*{A Clock in Stars?}

The stable model-inferred ages of $\mu$ Her and HD 49385 \citep{DM11} indicate that asteroseismic modelling can somehow overcome the model-dependence of stellar ages on chemical compositions and the convection parameters. Compared with the main-sequence stars, the only additional constraints for $\mu$ Her are the $\ell$ = 1 mixed modes, which correlates with the properties of the core. This suggests that a `clock' feature exists deep inside subgiants.
%However, it should be noted that the measurements of \textbf{effective temperature} ought to be primary condition for seismic modelling to give the constant age. Because the core significantly changes through the subgiant phase, without proper constraints of the atmosphere, a particular condition in the core could be satisfied by many of evolutionary tracks at different locations on the HR diagram. We hence only selected within a strict area on the HR diagram within $T_{\rm{eff}}$ = 5560$\pm$80 from our grid models. 

\begin{figure}
	% To include a figure from a file named example.*
	% Allowable file formats are eps or ps if compiling using latex
	% or pdf, png, jpg if compiling using pdflatex
  \begin{tabular}{c}
  \includegraphics[width=\columnwidth]{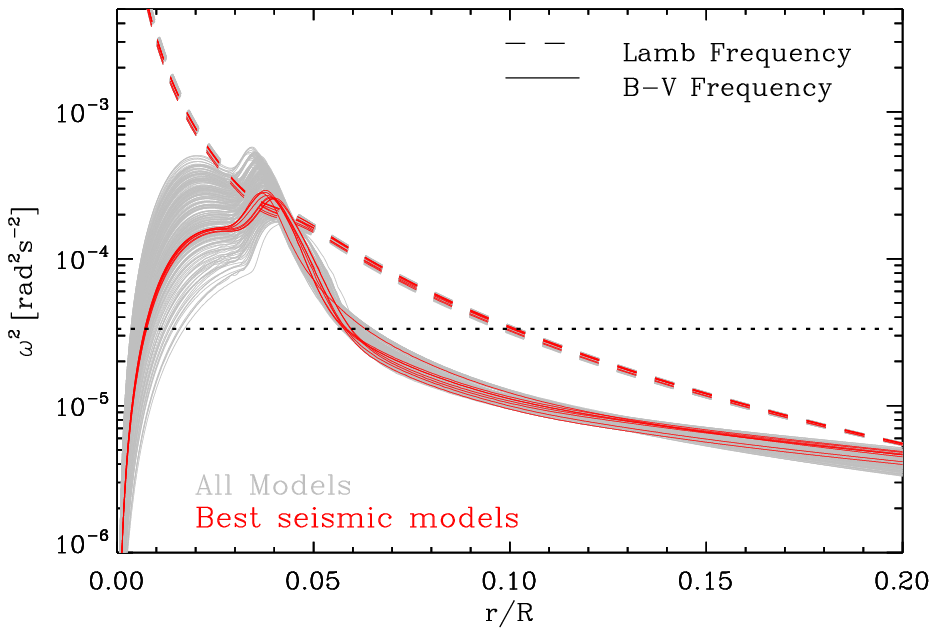}\\
  \includegraphics[width=\columnwidth]{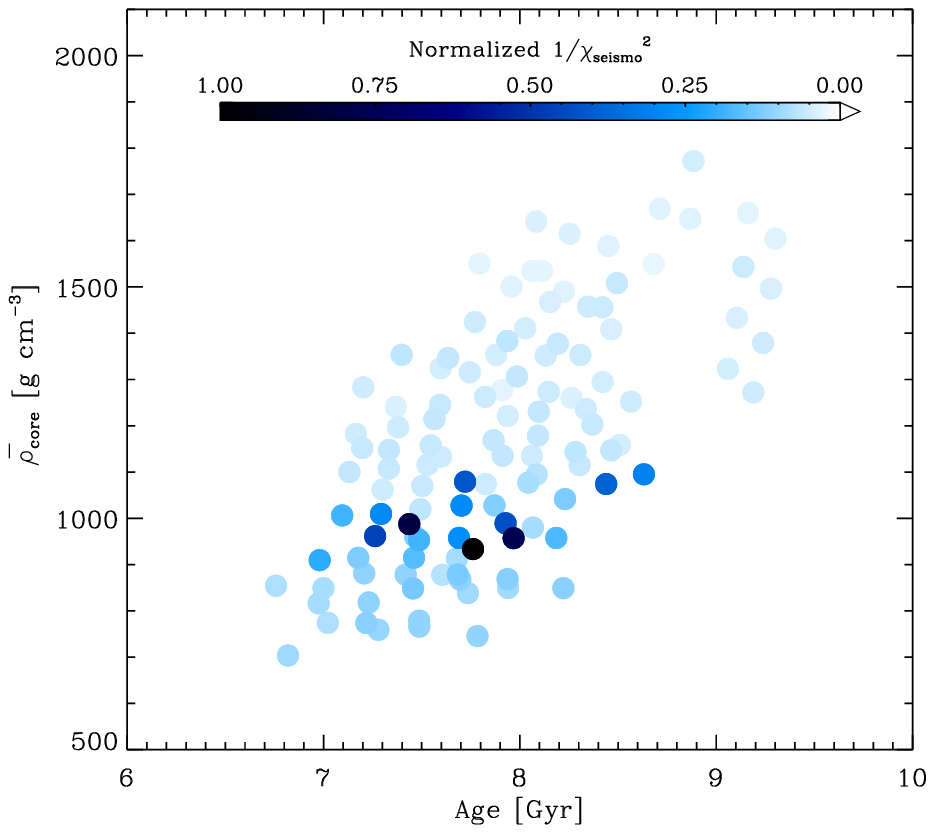} \\
  \end{tabular}
   \caption{ Top: the profiles of the $Brunt-V\ddot{a}is\ddot{a}l\ddot{a}$ frequency (solid lines) and $\ell$ = 1 $Lamb$ frequency (dashed lines) of the models within a strict limit of $T_{\rm{eff}}$ = 5560$\pm$80 K. Grey and red lines represent all grid models and the best seismic models (7 models in Table \ref{tab:topmodels} with $\chi^2_{\rm{seismo}}$ $\lesssim$ 25)). The dotted line shows the frequency of the bumped modes detected in $\mu$ Her.   
Bottom: the mean density of the helium core (where hydrogen fraction < 0.001) of the same models as a function of age. The colour indicates the normalised $1/\chi^2_{\rm{seismo}}$ (the darker colour, the better fit to the data). 
    \label{fig:clock}}
\end{figure}

As pointed out by \citet{DM11}, the frequencies of mixed modes of subgiants depend on the profile of the $Brunt-V\ddot{a}is\ddot{a}l\ddot{a}$ frequency ($N$) in the stellar core. Following this point, we plot the profiles of $N^2$ of all computed models in the top panel of Figure \ref{fig:clock} shown in grey. The profiles of the $Lamb$ frequency for $\ell$ = 1, which affects the acoustic modes, are also shown for comparison (dashed). Comparing the best seismic models shown in red (7 models in Table \ref{tab:topmodels} with $\chi^2_{\rm{seismo}}$ $\lesssim$ 25) illustrates that the $Lamb$ frequencies at a given depth do not show much spread, but significant differences in $N^2$ can be seen in the helium core ($r/R$ $\lesssim$ 0.04). This shows that the $Brunt-V\ddot{a}is\ddot{a}l\ddot{a}$ frequency in the core is the key to the `clock'. At the post-main-sequence phase, the $Brunt-V\ddot{a}is\ddot{a}l\ddot{a}$ frequency is mainly determined by the value of the central density because the core is almost isothermal \citep{DM11}. We hence plotted the mean densities of the helium cores (where the hydrogen fraction is below 0.001) of these models against stellar ages at the bottom of Figure \ref{fig:clock}. We see that our best-fitting models (indicated by the darker colour code) all have similarly dense cores, although their mass varies by 15\%. Thus, we find a clear correlation between the core density and the age. 

\section{Conclusions}
In this work, we used asteroseismic modelling for $\mu$ Her, which is a solar-like star in the subgiant stage, to constrain its stellar parameters. Compared with main-sequence stars, the appearance of the p-g coupled modes gives additional constraints on the stellar core, which significantly improves the accuracy of the modelling results. We then analysed the degeneracy of model parameters and found the age of $\mu$ Her to be weakly dependent on the mixing-length parameter and the chemical composition. The conclusions are summarised here:       
\begin{itemize}
\item The mass and age given by asteroseismic modelling for the star $\mu$ Her are 1.10$^{+0.11}_{-0.06}$ M$_{\odot}$ and 7.55$^{+0.96}_{-0.79}$ Gyr, which agrees well with the previous results using a different stellar evolutionary code and different input physics \citep{G17}. A similar modest sensitivity was found in the previous study of HD 49385 by \citet{DM11}, whose asteroseismic modelling also inferred very similar stellar masses and ages with
different input physics.
\item The initial helium abundance of the star was determined to be $\sim$0.28, suggesting a ratio of $\Delta Y/\Delta Z$ about 1.3 for $\mu$ Her, which is close to the solar value (1.50).
\item The inferred mixing-length parameter is $\sim$1.7, suggesting a $\sim$10\% smaller $\alpha_{\rm{MLT}}$ for the star than the Sun ($\alpha_{\rm{MLT, \odot}}$ = 1.9).    
\item Similar to the results in previous modelling efforts of other subgiants, the change in mixing length parameter does not affect the estimates of the mass and the age systematically.  
\item Asteroseismic modelling lifted the degeneracy between the age and chemical composition as well as the mixing-length parameter for the star. With a typical observed uncertainty of metallicity and an assumption of the initial helium (0.27 - 0.30) based on a standard helium and heavy element enrichment law, precise and accurate stellar ages ($\sigma$ <10\%) can be expected for stars similar to $\mu$ Her when using asteroseismology. Further studies on other subgiants are required to fully confirm this.
\item The mean density of the core, which affects the mixed modes of $\mu$ Her, seems to be a `clock' that reveals its age. 
%Within constraints on the surface ($T_{\rm{eff}}$, $\log g$, etc.),      
\end{itemize}
%\clearpage

\section*{Acknowledgements}
The research was supported by supported by an Australian Research Council DP grant DP150104667 awarded to JBH and TRB, the ASTERISK project (ASTERoseismic Investigations with SONG and {\em Kepler}) funded by the European Research Council (Grant agreement no.: 267864), Grants 11503039, 11427901, 11273007, and 10933002 from the National Natural Science Foundation of China.
Funding for the Stellar Astrophysics Centre is provided by The Danish National Research Foundation (Grant DNRF106). 

%%%%%%%%%%%%%%%%%%%%%%%%%%%%%%%%%%%%%%%%%%%%%%%%%%

%%%%%%%%%%%%%%%%%%%% REFERENCES %%%%%%%%%%%%%%%%%%

% The best way to enter references is to use BibTeX:

%\bibliographystyle{mnras}
%\bibliography{example} % if your bibtex file is called example.bib

% Alternatively you could enter them by hand, like this:
% This method is tedious and prone to error if you have lots of references

%%%%%%%%%%%%%%%%%%%%%%%%%%%%%%%%%%%%%%%%%%%%%%%%%%

%%%%%%%%%%%%%%%%% APPENDICES %%%%%%%%%%%%%%%%%%%%%

%\appendix

%\section{Some extra material}

%%%%%%%%%%%%%%%%%%%%%%%%%%%%%%%%%%%%%%%%%%%%%%%%%%

% Don't change these lines
\bsp	% typesetting comment
\label{lastpage}
\end{document}